\documentclass[pre,superscriptaddress,twocolumn,a4paper,showpacs]{revtex4-1}
\usepackage{graphics}
\usepackage{amssymb}
\usepackage{wasysym}
\usepackage{latexsym}
\usepackage{graphicx}
\usepackage{epsfig}
\usepackage{subfigure}
\begin{document}

\title{Policies for allocation of information in task-oriented groups: elitism and egalitarianism outperform welfarism}  

\author{Sandro M. Reia}
\affiliation{Instituto de F\'{\i}sica de S\~ao Carlos,
  Universidade de S\~ao Paulo,
  Caixa Postal 369, 13560-970 S\~ao Carlos, S\~ao Paulo, Brazil}

\author{Paulo F. Gomes}
\affiliation{Instituto de F\'{\i}sica de S\~ao Carlos,
  Universidade de S\~ao Paulo,
  Caixa Postal 369, 13560-970 S\~ao Carlos, S\~ao Paulo, Brazil}
\affiliation{Instituto de Ci\^encias Exatas e Tecnol\'ogicas, Universidade Federal de Goi\'as, 
75801-615  Jata\'{\i}, Goi\'as, Brazil }
  
  \author{Jos\'e F.  Fontanari}
\affiliation{Instituto de F\'{\i}sica de S\~ao Carlos,
  Universidade de S\~ao Paulo,
  Caixa Postal 369, 13560-970 S\~ao Carlos, S\~ao Paulo, Brazil}

\begin{abstract}
  Communication or influence  networks are probably the most controllable of all factors  that are known to  impact on the problem-solving capability of task-forces. In the case  connections
 are costly, it is necessary to implement a policy to allocate  them to the individuals.   
  Here we use an agent-based model to  study how distinct allocation policies affect the  performance of a group of agents
whose task is to find the global maxima of  NK fitness landscapes. Agents cooperate by broadcasting  messages  informing on their fitness and use this information to imitate  the fittest agent in their influence neighborhoods. The larger the influence neighborhood of an agent, the more links, and hence information, the agent receives.  We find that 
the elitist policy in which  agents with above-average fitness  have their influence neighborhoods amplified, whereas agents  with below-average fitness  have theirs deflated, is optimal  for smooth landscapes, provided the group size is not too small. For rugged landscapes, however, the elitist policy  can  perform very poorly  for certain group sizes. In addition,  we find that the egalitarian policy, in which  the size of the influence neighborhood is the same for all agents, is optimal for both smooth and rugged landscapes in the case of small groups.
The welfarist policy, in which the actions of the elitist policy are reversed, is always suboptimal, i.e., depending on the group size it is outperformed by either the elitist or the egalitarian policies.  
%
\end{abstract}

\maketitle

\section{Introduction}\label{sec:intro}

Solving real-world problems typically  entails overcoming obstacles  that are beyond the capabilities of a single person,  thus requiring   the cooperative and coordinated effort  of  individuals organized in task-oriented groups or  epistemic communities, broadly viewed 
as  social systems consisting of producers of knowledge \cite{Reijula_19,Kitcher_93}.
Information in these systems flows between individuals via social contacts and, in the cooperative problem-solving  setting, a key process is imitative learning  as  expressed in this quote by Bloom ``Imitative learning acts like a synapse, allowing information to leap the gap from one creature to another'' \cite{Bloom_01}. In  the context of epistemic communities, imitative learning is known as exploitation  since it corresponds to the search strategy  in which the agent  borrows  known solutions from its network of influencers. This contrasts with exploration,  which is seen as the development of novel solutions by the agent \cite{Lazer_07,Goldstone_08}.

 Here we  assume that  the amount of information an agent can access (i.e., the number of its potential influencers)  has a social cost (e.g., it is publicly funded) and we examine the influence of different policies of allocation of resources to the agents on the efficiency  of the group to complete  the task.   These policies are based on the quality of the partial solutions  the agents offer to solve the  task.
To address this problem we  use an agent-based  model where  the agents perform individual trial-and-test searches to probe a fitness  landscape (exploration)  and  imitate  a model agent -- the best performing agent in their influence neighborhood at the trial (exploitation) \cite{Fontanari_14,Fontanari_15,Reia_19}.  We consider a scenario where   the agents are fixed at the nodes of a random geometric graph \cite{Gilbert_61} and  can interact with each other  if the distance between them is less   than a prespecified threshold.  In addition, the agents vary their radiuses of interaction following a prescription or policy that depends only  on their  (relative) fitness. This feature   results in a time-dependent,  adaptive directed network that links the agents to their influencers.

The task of the agents is to find the global maxima of smooth and rugged fitness landscapes generated with the NK model \cite{Kauffman_87}.  The performance of the group is measured by the properly scaled number of trials (or time) required to find those maxima. In particular,  the collective search ends when one of the agents finds the global maximum. 
The amount of information allocated to an agent is determined by the number of agents in  its influence neighborhood, which equals the number of incoming  connections,  and here we consider three distinct  policies.  The elitist policy in which the agents with above-average fitness amplify their influence neighborhoods whereas the agents with below-average fitness shrink  theirs; the welfarist policy in which those actions are reversed; and the egalitarian policy in which the size of the influence neighborhood is the same for all agents, regardless of their fitness.
Since the fitness of the agents  change as they explore the state space of the fitness landscape, so do their influence neighborhoods,  resulting in adaptive directed  communication networks, which  we  characterize   through the number and size of their strongly connected components.
 
 The   scenario considered here bears a resemblance to  the predicament that funding agencies  and governments face when  allocating  resources among alternative competing research programs aiming at solving the same problem \cite{Reijula_19,Kitcher_93}. In fact,  the priority rule for allocating credit in science  by which  only the first person to a discovery gets the recognition supports our decision to  halt the search the first time  an agent finds the global maximum: once a result has been discovered, no value to the collective is produced by discovering it again
  \cite{Merton_73,Streven_03}. In that sense, welfarism   is not a so far-fetched policy for resource allocation since it is arguably a sensible strategy from  the perspective of the collective good to foment  agents with below-average performances in order to maintain a diversity of  approaches to  unsolved problems.  In addition,  similar agent-based models have also been used to study welfare in the context of human cooperation  \cite{Perc_17}.

 We find that for both smooth and rugged landscapes the welfarist policy is always suboptimal, i.e., it is outperformed either by the elitist or by the egalitarian policies. In addition, for small group sizes the egalitarian policy always yields  the optimal performance. Except for small groups, the elitist policy is the optimal choice  in the case of  the smooth landscapes without local maxima, but a too large amplification of the
  influence neighborhoods of the agents with above-average fitness may seriously harm the performance of groups of intermediate size 
 in the case of rugged landscapes.  As expected, high-fitness outliers in the initial randomly generated group are very likely  to win  the search (i.e., to find the global maximum first)  under  the elitist policy. Most surprisingly, however,  is  the finding that even in a situation of strong welfare,  the high-fitness outliers of the initial generation are  still more likely to become winners, so the welfarist  policy cannot reverse the random initial fitness inequality.

The rest of the paper is organized as follows. In Section \ref{sec:NK},  we present a brief description of the NK model of  
rugged fitness landscapes.  In Section \ref{sec:model},  we describe the rules for setting up  the influence neighborhoods  of the agents as well as the implementation of the imitative  learning search to explore the state spaces of  fitness landscapes. In Section \ref{sec:res}, we study the performance of this search for both smooth and rugged landscapes focusing on the effects of the group size and of the policies of allocation of information.
Finally, Section \ref{sec:conc} is reserved for our concluding remarks.

\section{NK-fitness landscapes}\label{sec:NK}

The NK model \cite{Kauffman_87} is a computational implementation  of fitness landscapes that 
has been extensively used to study optimization problems in population genetics, developmental biology and protein folding \cite{Kauffman_95}.  It was   introduced originally   to model the adaptive evolution process as walks on rugged fitness landscapes and its  main advantage, which led to its widespread use in complexity science,  is the possibility of tuning  the ruggedness of the landscape by changing the two integer parameters that give the model its name,  namely, $N$ and $K$. More pointedly,
the NK landscape is defined in the space of binary strings of length $N$ and so the parameter $N$ determines the size of the state space, $2^N$.  The other parameter  $K =0, \ldots, N-1$  determines the range of the epistatic interactions among the bits of the binary string and  influences strongly the number of local maxima on the landscape. In time,  two bits are said to be epistatic whenever the combined effects of their contributions to the fitness of  the binary string  are not  merely additive \cite{Kauffman_87}.
In particular,
for $K=0$ the corresponding (smooth) landscape has one single maximum whereas for $K=N-1$, the (uncorrelated) landscape  has on the average  $2^N/\left ( N + 1 \right)$ maxima with respect to single bit flips \cite{Kaul_06}. 

In the NK model, each string  $\mathbf{x} = \left ( x_1, x_2, \ldots,x_N \right )$ with
$x_i = 0,1$ has  a fitness value $\Phi \left ( \mathbf{x}  \right ) $ that is given by  the average  of the contributions  of each 
component $i$ in the string, i.e.,
\begin{equation}\label{Phi}
\Phi \left ( \mathbf{x}  \right ) = \frac{1}{N} \sum_{i=1}^N \phi_i \left (  \mathbf{x}  \right ) ,
\end{equation}
where $ \phi_i$ is the contribution of component $i$ to the  fitness of string $ \mathbf{x} $. It is assumed that $ \phi_i$ depends on the state $x_i$  as well as on the states of the $K$ right neighbors of $i$, i.e., $\phi_i = \phi_i \left ( x_i, x_{i+1}, \ldots, x_{i+K} \right )$ with the arithmetic in the subscripts done modulo $N$.  It is assumed, in addition, that  the functions $\phi_i$ are $N$ distinct real-valued functions on $\left \{ 0,1 \right \}^{K+1}$ and, as usual,  we assign to each $ \phi_i$ a uniformly distributed random number  in the unit interval \cite{Kauffman_87}. Because of the randomness of $\phi_i$, we can guarantee that  $\Phi   \in \left ( 0, 1 \right )$ has a unique global maximum and that different strings have different fitness values. 

For $K=0$ there are no local maxima and the sole maximum of $\Phi$ is easily located by picking for each component $i$ the state $x_i = 0$ if  $\phi_i \left ( 0 \right ) >  \phi_i \left ( 1 \right )$ or the state  $x_i = 1$, otherwise. However, for $K>0$ finding the global maximum of the NK model is a NP-complete problem \cite{Solow_00}, which  means that the time required to solve the problem using any currently known deterministic algorithm increases exponentially fast with the length $N$ of the strings.
The increase of the parameter $K$ from $0$ to $N-1$  decreases the correlation between the fitness of neighboring strings 
(i.e., strings that differ at a single component) in the state space and for $K=N-1$, those fitness values are  uncorrelated \cite{Hordijk_19}.
 
Since the functions  $ \phi_i$ in eq.\ (\ref{Phi}) are random, the ruggedness measures (e.g., the number of local maxima) of a particular realization of a NK landscape are not fixed by the choice of the parameters $N$ and $K>0$. In fact, those measures  can vary considerably between landscapes with the same values of those parameters  \cite{Kauffman_87}, which implies that the  performances of  search  heuristics that rely on the  local correlations of the fitness landscape  will depend on the particular realization of that landscape. Thus,  in order to highlight the role of the parameters  that are relevant to our goal of exploring the effects of  the policies of allocation of  information to the agents on group performance, here we compare the performance  of the  groups for the same realizations of the NK fitness landscapes.
In particular, we consider two types of NK landscapes: smooth landscapes with  $N=12$ and $K=0$,  and   rugged landscapes with $N=12$ and $K=4$. For fixed $N$,  all NK landscapes with $K=0$ are equivalent and so  we can  consider a single realization of the  smooth NK landscapes without lack of generality.  For $K=4$, however,  we must average the group performance over an ensemble of landscapes in order to obtain statistical  meaningful results. To guarantee  that the groups solve the same tasks we generate and store  a set of 30 landscape realizations with parameters $N=12$ and $K=4$ so the same landscape realizations are used for different  parameters of the  imitative learning search. The minimum and the maximum number of maxima in the landscape realizations of our ensemble are $31$ and $56$, respectively, whereas the mean number of maxima   is $46.5$. 

\section{Imitative learning search}\label{sec:model}

We consider a system of $M$ agents placed  in a   square box of linear size $L$ with periodic boundary conditions. In the initial configuration, the coordinates $x$ and $y$ of each agent are chosen  randomly and uniformly over the  length $L$. The density of agents
$\rho = M/L^2$,  which we fix to $\rho = 1$ throughout the paper, yields the relevant spatial scale to measure the distance between  agents on the square box.  In fact, since the effective  area of an agent is $1/\rho$, the  quantity $d_0 = 1/\sqrt{\rho}$ can be viewed as the  linear size or, for short, the size of an agent and it will be our standard  to measure all distances in our study.   We note that the fixed value of  the density $\rho$ is inconsequential, provided that we use $d_0$ as the standard for measuring distances in the square box.  Each agent is represented by a binary string  of length $N$, whose bits  are  initially   drawn at  random with equal probability for $0$ and $1$, so that each agent has an associated fitness value $\Phi_k$ with $k=1, \ldots, M$.  The fitness of the agents  change with time as
they explore the NK-fitness landscape aiming at finding its global maximum  by flipping bits  following the rules of the imitative learning search \cite{Fontanari_15} as will be described next. Henceforth we will use the terms agent and string interchangeably.

The influence neighborhood  of agent $k$ is comprised of all the agents  located inside  the circle of radius $d_k$ centered at the spatial coordinates of  agent $k$. It is among those agents   that  agent $k$ will select a model to  imitate. Here we  consider the prescription
\begin{equation}\label{dk}
d_k = d_0 \exp \left [ \alpha \left ( \Phi_k / \bar{\Phi} - 1 \right ) \right ]
\end{equation}
 where $\bar{\Phi} =   \sum_{k=1}^M \Phi_k  /M$ is the mean fitness of the group at time $t$ and, for the sake of clarity, we have omitted the dependence on $t$ of the quantities  $d_k$, $\Phi_k $ and $\bar{\Phi} $. The parameter $\alpha$ determines the radius of the influence neighborhood of each agent according to its relative fitness. For $\alpha > 0$, agents with fitness higher than average have a large influence neighborhood,  i.e., they can see and eventually copy more agents in the group, whereas  the agents with fitness lower than average have their influence neighborhoods downsized and are likely to become isolated for large $\alpha >0$.  We refer to this choice as the elitist policy, since the high-fitness agents have more opportunities to  further  improve their fitness through imitation.
 For $\alpha <0$ the situation is reversed so that the lower-than-average fitness agents have their neighborhoods amplified and the above-than-average fitness agents have theirs curtailed. We refer to this choice as the welfarist policy.  The case $\alpha =0$, where the sizes of the influence neighborhoods are the same for all agents, corresponds to an egalitarian policy.
 
\begin{figure} 
\centering  
 \includegraphics[width=0.48\textwidth]{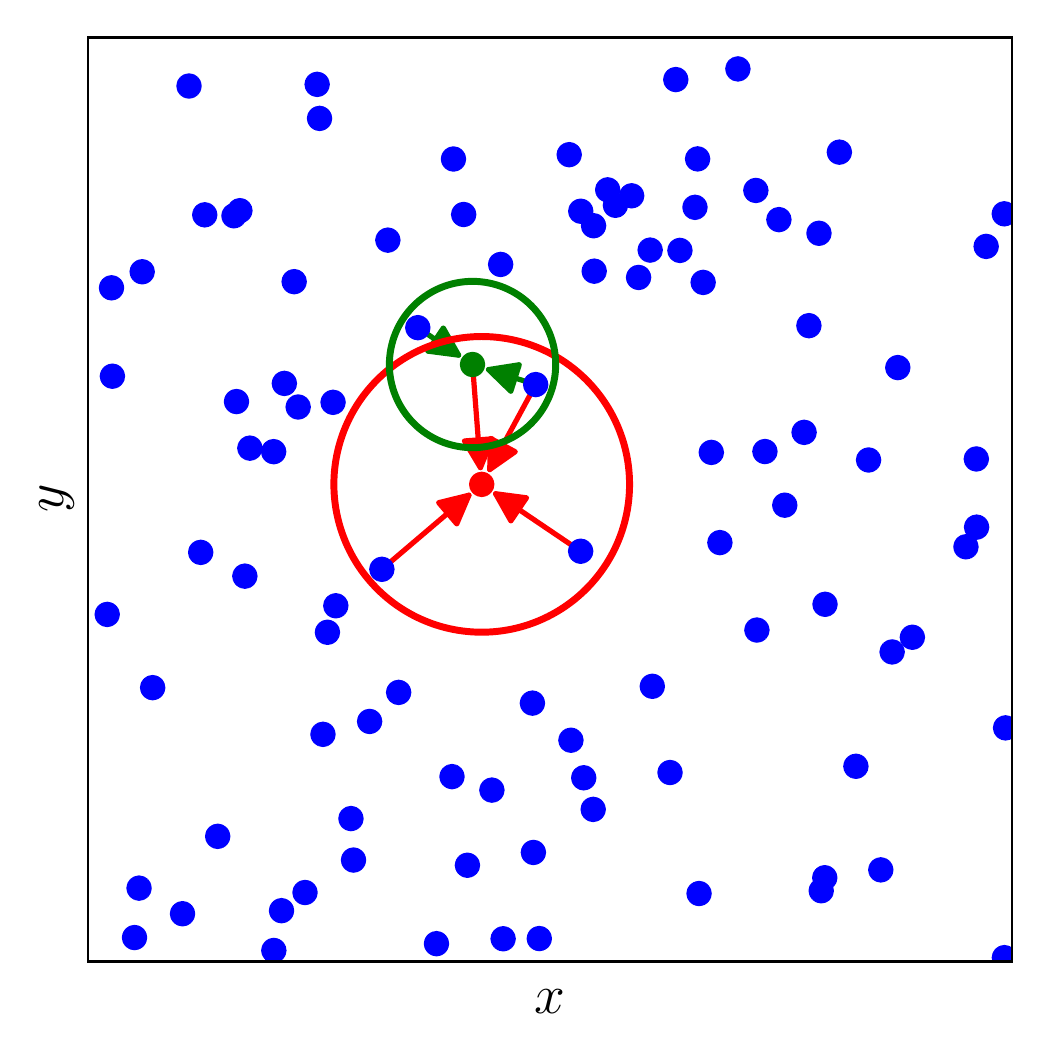}  
\caption{Snapshot of a system of $M=100$ agents, represented by the bullets, distributed randomly in a square box with density $\rho = 1$. The influence neighborhoods are shown for two  selected agents  only  and are highlighted by  circles of different colors.  This example shows that the resulting influence network  is a  directed network.
 }  
\label{fig:1}  
\end{figure}

 This scenario is illustrated in Fig.\ \ref{fig:1} that shows a snapshot of  a system of  $M=100$ agents in the square box.  Henceforth we refer to the network created  by the union of the  influence neighborhoods  of all agents as the influence network. This directed network reduces to   the classic undirected random geometric graph  for $\alpha =0$. The random geometric graph was originally introduced to model wireless communication networks \cite{Gilbert_61} and it  was recently  used  as a face-to-face network  in the  modeling of the  dynamics of human interactions \cite{Starnini_13} as well as in the study of the effects of mobility on cooperative processes  \cite{Gomes_19}. 

The dynamics begins with the selection of a target agent at random, say agent $k$,  at time $t=0$ and  proceeds as follows. A circle of radius $d_k$  is drawn around the target agent so that its influence neighborhood is determined (see Fig.\ \ref{fig:1}).  If the influence neighborhood is empty, i.e., there is no agent within a distance $d_k$ from agent $k$, or all agents in the
influence neighborhood   have fitness lower than  or equal to the fitness of the agent $k$, then this agent  simply flips a  bit  at random. We recall that due to the nature of the NK-fitness landscape -- the fitness are real-valued random variables -- two agents that have the same fitness must be identical (clones). If there are agents with fitness higher than the fitness of the target agent in its influence neighborhood, there are two possibilities of action. The first action, which happens with probability $1-p$,   consists of flipping a  bit  at random of the target string as before. Through the repeated application of this action,  the agents can produce all  the $2^N$  binary strings starting from any arbitrary string, which guarantees that the global maximum will  eventually  be reached for $p < 1$.  The second action, which happens with probability $p$, is the  imitation  of a model string, which is the string  of highest fitness  in the  influence  neighborhood of the target agent. In this case, the  model and the target  strings are compared  and the different bits are singled out.  Then the target agent selects at random one of the distinct bits and flips it so that this bit is now the same in both strings.   Hence,  imitation  results in  the increase of the similarity between the target and the model agents, which may not necessarily lead to  an increase of the fitness of the target agent.   

The parameter $p \in \left [0,1 \right ]$ is the imitation probability, which we assume is  the same for all agents (see \cite{Fontanari_16} for the relaxation of this assumption).  The case $p=0$ corresponds to the baseline situation where  the $M$ agents explore the state space independently  of each other.  The case $p=1$ corresponds to the situation where only the model strings explore the state space through  random bit flips, whereas the other strings simply follow the models in their influence neighborhoods.  The imitation procedure described above was borrowed from the incremental assimilation mechanism used to study the influence of external media  \cite{Shibanai_01,Avella_10,Peres_11} in   Axelrod's model  of social influence \cite{Axelrod_97}.  This is the main feature that distinguishes our model from previous exploration (random bit flips) and  exploitation (copy of fittest agent) models in which the copy mechanism is non-incremental so that the target agent is replaced by the model agent \cite{Lazer_07}. As expected, this non-incremental mechanism  may permanently trap the search in the local maxima  of the fitness landscape.

After the target agent is updated, which means that exactly one bit of  its string is flipped,   we increment the time $t$ by the quantity $\Delta t = 1/M$.   Then another agent is selected at random and the procedure described above is repeated. We note that during the increment from $t$ to $t+1$, exactly  $M$  string operations are performed, though not necessarily by $M$ distinct agents. 
The  search ends when one of the agents finds the global maximum and we denote by $t^*$  the halting time.  
Here we measure the efficiency of the  search by the total number of agent updates necessary to find the global maximum (i.e., $Mt^*$), which is essentially  the computational cost of the search.  Since $t^*$ typically scales    with the size of the solution space $2^N$, it is convenient to present the results in terms of the rescaled computational cost,  defined as 
\begin{equation}\label{C}
C = M t^*/2^N .
\end{equation}
In  the case of the independent search ($p=0$), the ruggedness of the landscape has no effect on the efficiency  of the search, which depends only on the length of the strings $N$ and on the group size $M$. In fact, in this case it can be shown that the mean rescaled computational cost is given by 
\begin{equation}\label{Cind}
\langle C \rangle = \frac{M}{ 2^N \left [ 1 - \left ( \lambda_N \right)^M \right ]},
\end{equation}
where  $\lambda_N$ is the second largest eigenvalue of a tridiagonal stochastic matrix $\mathbf{T}$ \cite{Fontanari_15}.  The  notation $\langle \ldots \rangle$  stands for the average over independent searches on the same landscape.
Notice that  $\langle t^*_M \rangle = 1/\left [ 1 - \left ( \lambda_N \right)^M \right ]$ is the expected number of trials for a group of $M$ independent agents to find the global maximum. 
In particular, for $N=12$ we have  $\lambda_{12} \approx 0.99978$ and  $\langle t^*_1 \rangle \approx  4545$. Since 
$\left ( \lambda_{12} \right)^M \approx e^{- M \left ( 1 - \lambda_{12} \right )}$ we have 
$\langle C \rangle \approx \langle t^*_1 \rangle / 2^{12} \approx 1.11$ for $M \ll \langle t^*_1 \rangle$  and
$\langle C \rangle \approx M/ 2^{12} $ for $M \gg \langle t^*_1 \rangle$.

The assessment of the performance of the imitative learning search is done by comparing its mean computational cost with the   cost of the independent search, which is  approximated very well by the constant  $\langle C \rangle  \approx 1.11$ for the typical group sizes $M$ considered in the paper.

As pointed out before, the performance of the imitative search is  measured by the mean  computational cost $\langle C \rangle $, which  is estimated by averaging the computational cost defined by eq.\ (\ref{C}) over $10^4$  searches on the same  landscape realization. For the
rugged landscapes, the resulting cost is further averaged over the set of 30 landscape realizations.  Since  our main concern  is the effect of the resource allocation policies   on  group performance, we will fix  the imitation probability to $p=0.5$ and  vary  the  group size   $M$ and  the parameter $\alpha$   that appears in eq.\ (\ref{dk}) and  determines the strength with which the elitist ($\alpha > 0$) and welfarist ($\alpha < 0$) policies are enforced, i.e., the value of $\alpha$ determines how the radius of the influence neighborhood of an agent is affected  by  its relative fitness.

\subsection{Smooth Landscape}

\begin{figure}
\centering
\includegraphics[width=0.48\textwidth]{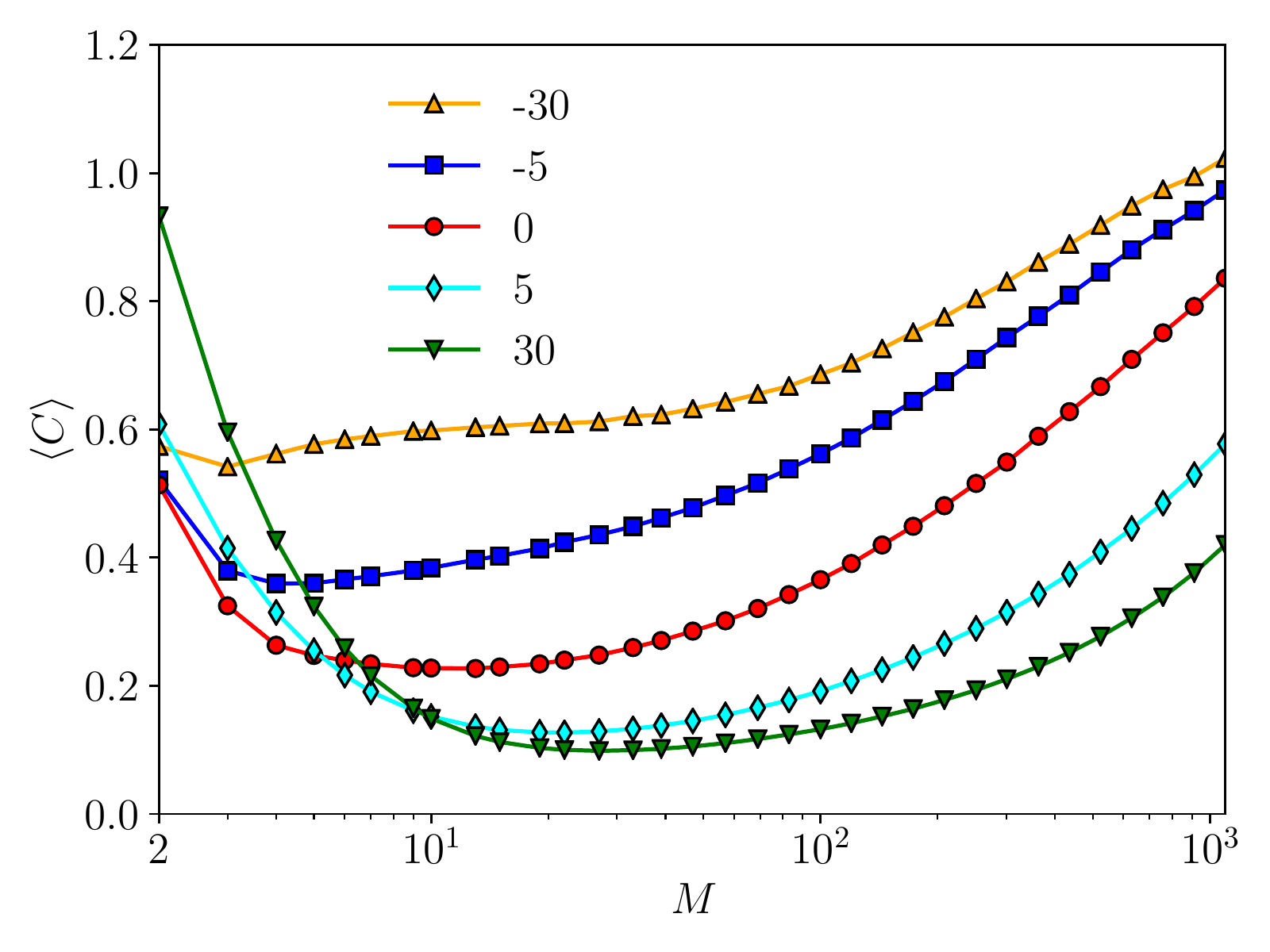}
\caption{Mean computational cost $\langle C \rangle $ as function of the group size $M$ for 
the imitative search on a smooth landscape. 
The imitation probability is $p=0.5$ and  the strength with which the  information allocation policies are enforced is
 $\alpha = -30, -5, 0, 5, 30$ as indicated. 
 The parameters of the NK landscape are $N=12$ and $K=0$.  }
\label{fig:2}
\end{figure}

\begin{figure}
\centering
\includegraphics[width=0.48\textwidth]{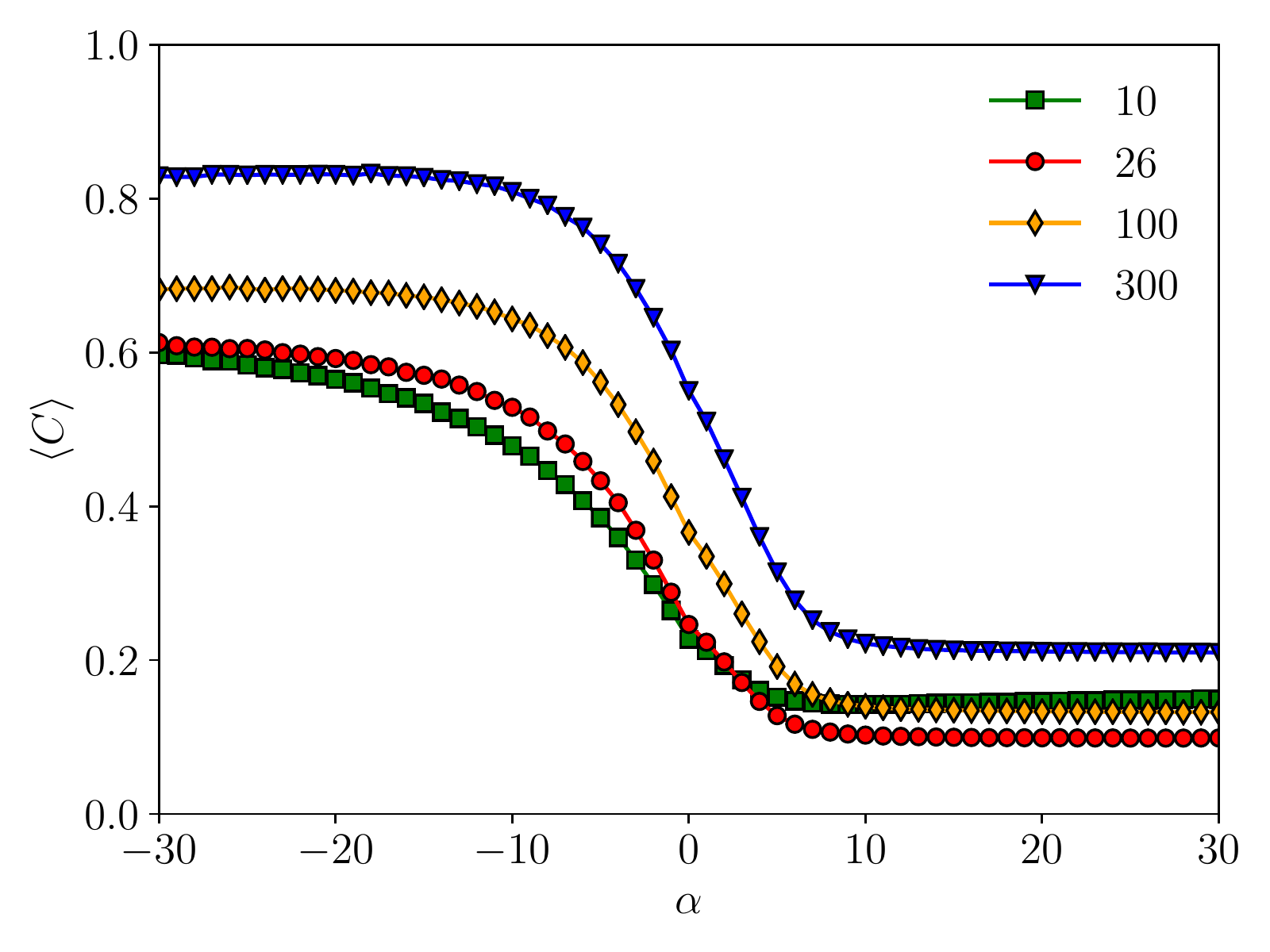}
\caption{Mean computational cost $\langle C \rangle $ as function of the strength $\alpha$  with which the  information allocation policies are enforced  for  group sizes $M=10,26, 100, 300$ as indicated. 
The imitation probability is $p=0.5$  and the parameters of the NK landscape are $N=12$ and $K=0$.  }
\label{fig:3}
\end{figure}

The  NK fitness landscape with $K=0$  is an additive landscape (i.e., the fitness of a string is given by the sum of the fitness of its components)  that exhibits a single maximum.  For the particular  realization we consider here, the   fitness of the maximum is  $\Phi^{\mbox{\small max}} = 0.559$,   whereas the average fitness of the landscape is  $\Phi^{\mbox{\small av}} = 0.415$. The  mean computational cost of the imitative search for a landscape with $N=12$ and $K=0$ is shown in Figs.\ \ref{fig:2} and \ref{fig:3}, where  the  independent variable is $M$  and $\alpha$, respectively. As already pointed out, these results are not dependent on the realization of the  smooth landscape.

It is convenient to begin the analysis of  Fig.\ \ref{fig:2} with the results for  $\alpha=30$, where we observe an initial decrease of 
 the computational cost  with increasing $M$ until  it reaches a minimum at the optimal group size  $M = 26$.  The subsequent increase of the cost for $M$ greater than  this optimum is probably due to the concentration of the strings in the vicinity of the model string and the consequent production of clones that end up reducing the efficiency of the search  \cite{Fontanari_15}.  The optimal group size decreases and the group performance degrades  with decreasing $\alpha$. For instance, for $\alpha = -30$, the mean computational cost reaches its minimum value  at $M=3$.

%
%

%
%

 Figure \ref{fig:3} shows  the computational cost against $\alpha$ and reveals more clearly the interesting result that  the group performance improves with increasing $\alpha$, provided that  $M$ is not too small. This means that for the imitative search on a smooth landscape it is advantageous to allow the   above-average fitness agents to enlarge their influence neighborhoods so they can inspect and eventually imitate  more agents in the group. This elitist policy increases the chances of improvement of the agents which already have a high fitness  and decreases those of the low-fitness agents, similarly to the so-called Matthew principle in which the rich get richer and the poor get poorer \cite{Merton_68}. The opposite, welfarist policy in which the below-average agents enlarge their neighborhoods (i.e., $\alpha < 0$) results in a much poorer performance, as illustrated in Figs.\ \ref{fig:2} and \ref{fig:3}.
 Interestingly, however, for small groups, say $M< 5$ in  Fig.\ \ref{fig:2}, the best performance is achieved for the egalitarian policy ($\alpha =0$) where the sizes of the influence neighborhoods are not dependent on the fitness of the  agents (i.e., $d_k = d_0$ for  $k=1,\ldots,M$). To conclude the analysis of the group performance, we stress that the imitative search always outperforms  the independent search for the smooth landscape.
 
\begin{figure}
\centering
 \subfigure{\includegraphics[width=0.48\textwidth]{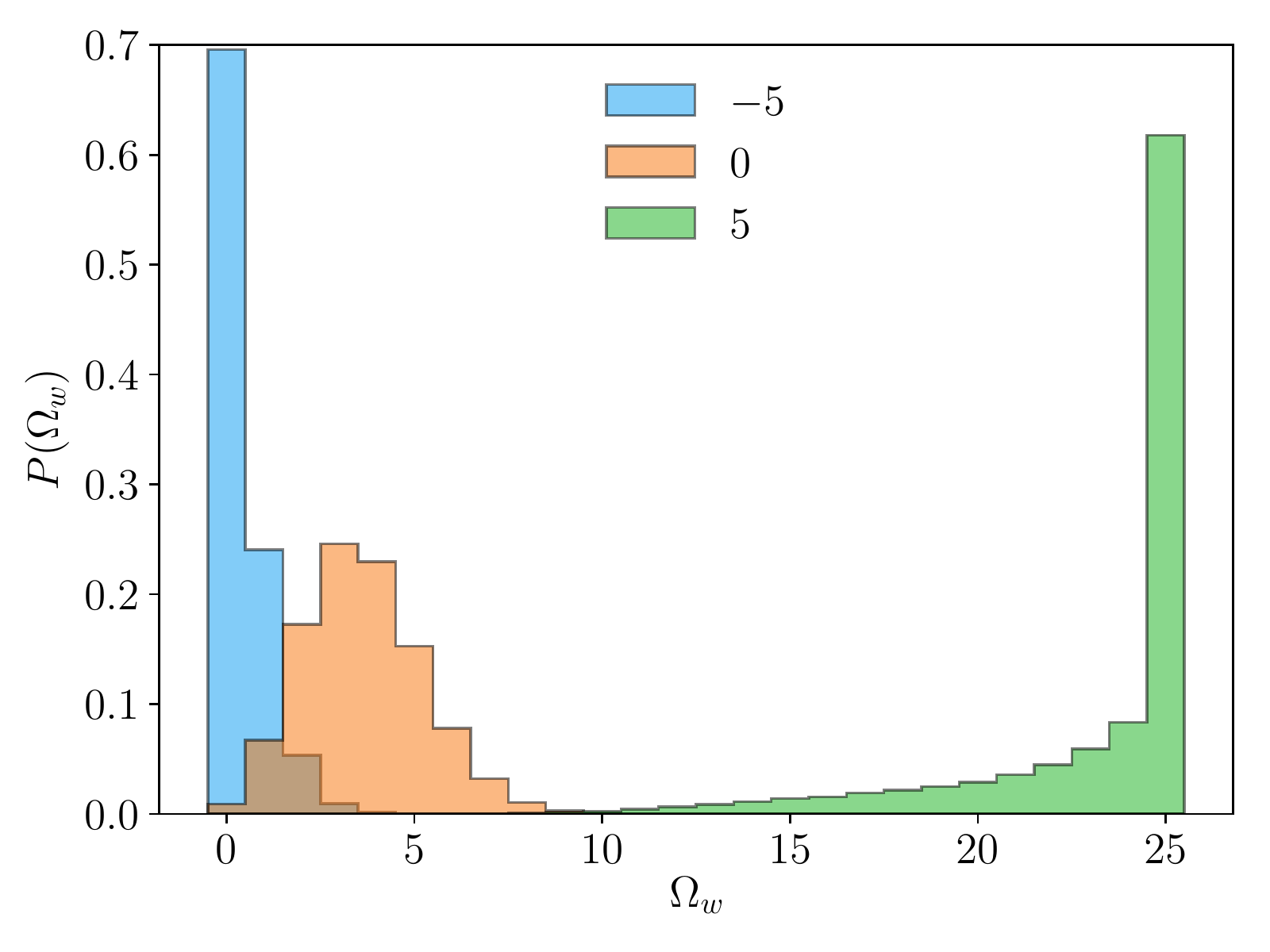}} \\
\subfigure{\includegraphics[width=0.48\textwidth]{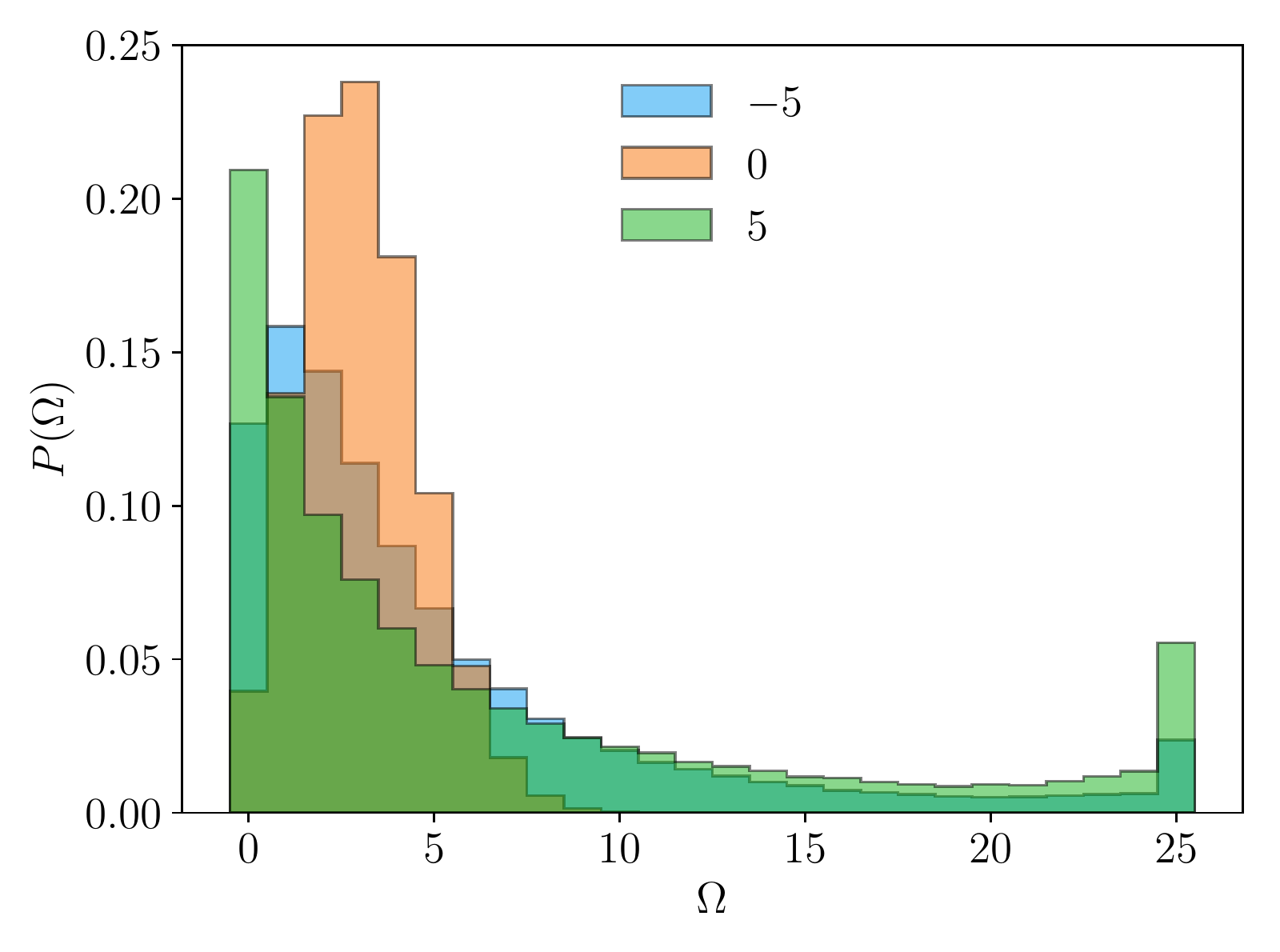}}
\caption{Probability distribution of the number of agents in the influence neighborhood of the winner (upper panel) and of 
a randomly selected agent (lower panel)  at the  instant just before the winner finds the global maximum for 
 $M=26$ and $\alpha =-5,0,5$   as indicated. 
The imitation probability is $p=0.5$ and the parameters of the NK landscape are $N=12$ and $K=0$.  }
\label{fig:4}
\end{figure}
 
It is instructive to look  into the characteristics of the agent that  found the global maximum and, con\-se\-quent\-ly, halted the search.  We refer to this agent as the winner. A useful quantity in this context is the distribution of probability of the number of agents in the winner's influence neighborhood, $\Omega_w = 0, \ldots, M-1$, at the instant just before it finds the global maximum, which we  show in the  upper panel of Fig.\ \ref{fig:4}. For the purpose of  comparison,  we show in the  lower panel of Fig.\ \ref{fig:4}  the distribution of probability of the number of agents in the influence neighborhood of a randomly selected agent at the same instant.  We  note that $P \left ( \Omega \right )$ gives effectively the distribution of the sizes of the influence neighborhoods at the trial just before the search halts.
For $\alpha < 0$, these results indicate that  the winner is very likely to be an isolated agent (i.e., $\Omega_w = 0$), as expected, though there are little more than 10\% of isolated agents when the search halts. In fact, since the winner must differ of exactly one bit from the global maximum just before it flips the discordant bit, its fitness must be high  and, consequently, its influence neighborhood must be small. The same reasoning applies for $\alpha >0$, so the winner is very likely to be connected to all the other agents (i.e., $\Omega_w = M-1$), although, in this case,  there are very few highly connected agents in the group as shown in the lower panel of  Fig.\ \ref{fig:4}.  Hence, regarding the sizes of their influence neighborhoods and provided an information  allocation policy is enforced (i.e., $\alpha \neq 0$), the winners are atypical agents just  before they find the global maximum.

\begin{figure}
\centering
\includegraphics[width=0.48\textwidth]{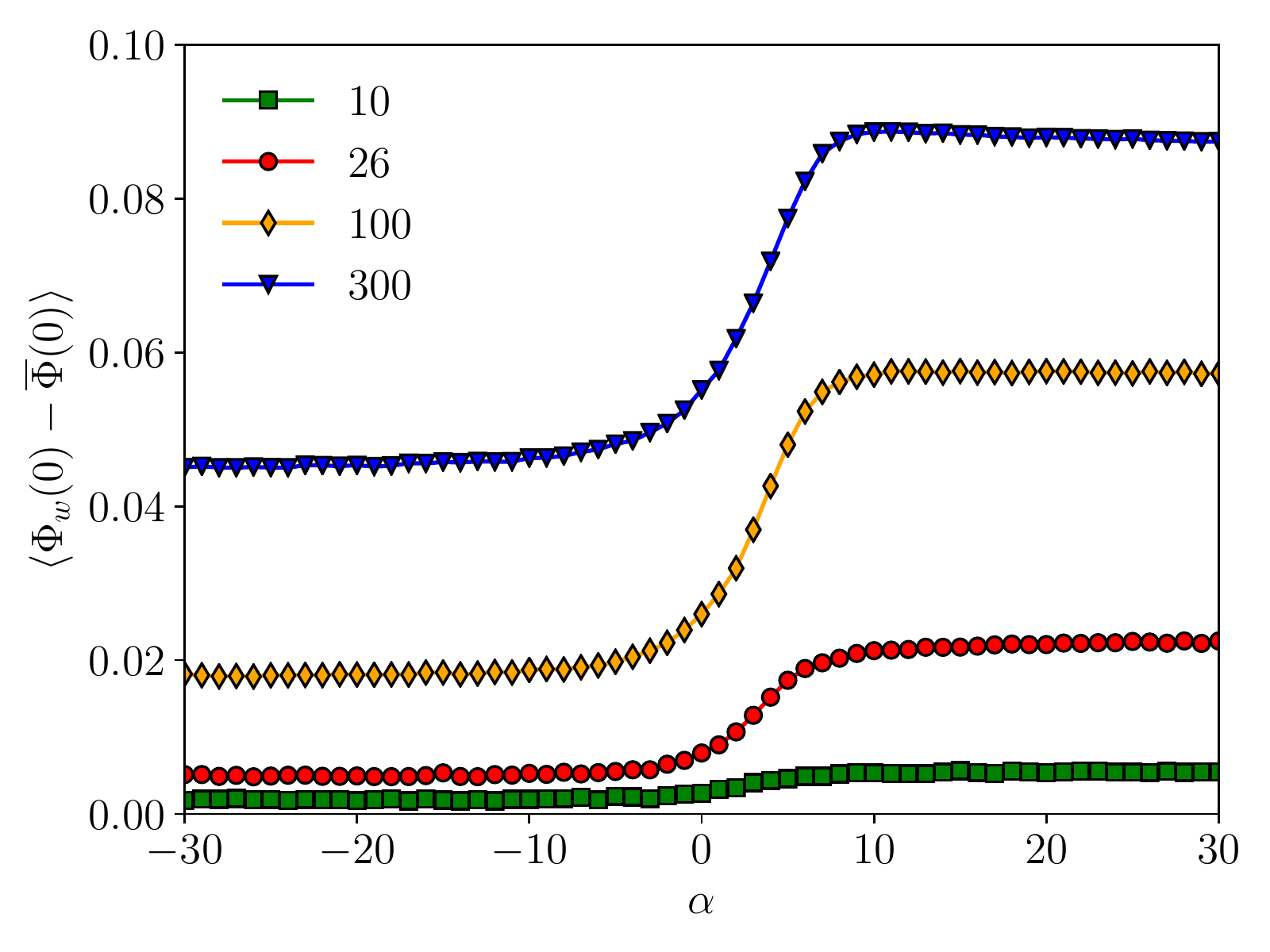}
\caption{Average difference between the initial fitness of  winner   and the  initial mean fitness of the group 
$\left  \langle \Phi_w \left ( 0 \right ) - \bar{\Phi} \left ( 0 \right ) \right \rangle$  as function 
of the strength $\alpha$  with which the  information allocation policies are enforced
 for  group sizes $M=10,26, 100, 300$ as indicated. 
The imitation probability is $p=0.5$  and the parameters of the NK landscape are $N=12$ and $K=0$.  }
\label{fig:5}
\end{figure}

 In order to investigate whether the winners had an edge in the initial random setup of the group, we calculate the difference between the fitness of the winners at time $t=0$, $ \Phi_w \left ( 0 \right ) $,  and the initial mean fitness of the group, $ \bar{\Phi} \left ( 0 \right )$, for each search. Then we average this difference over the $10^4$ searches and show the result
 $\left  \langle  \Phi_w \left ( 0 \right ) - \bar{\Phi} \left ( 0 \right ) \right \rangle $   in Fig.\ \ref{fig:5}.  Here the notation $\left  \langle \ldots \right \rangle$ represents an average over different searches and, in the case of rugged landscapes, over landscape realizations, whereas  $\bar{\ldots}$ stands for an average over the agents in the group. The effect of the group size $M$ can be understood by noting that for small $M$ the chances that an agent   -- an outlier-- is assigned a high fitness value at $t=0$   are very meager and so, in this case,  $ \Phi_w \left ( 0 \right ) $ does not differ  much from the  group average $ \bar{\Phi} \left ( 0 \right ) $. As $M$ increases, the  chances that an outlier appears increase and the result that  $\left  \langle \Phi_w \left ( 0 \right ) - \bar{\Phi} \left ( 0 \right ) \right \rangle  > 0$ indicates that those outliers are more likely to be the winners, regardless of the value of $\alpha$. The dependence of $\left  \langle \Phi_w \left ( 0 \right ) - \bar{\Phi} \left ( 0 \right ) \right \rangle $ on $\alpha$ for fixed $M$ is more instructive, as it shows that the elitist policy  ($\alpha > 0$) practically selects  the future winners already in the initial generation by
 allowing the fittest agent to reap all the benefits of imitative learning.  This is again an illustration of the Matthew principle in action.
 
\begin{figure}
\centering
 \subfigure{\includegraphics[width=0.48\textwidth]{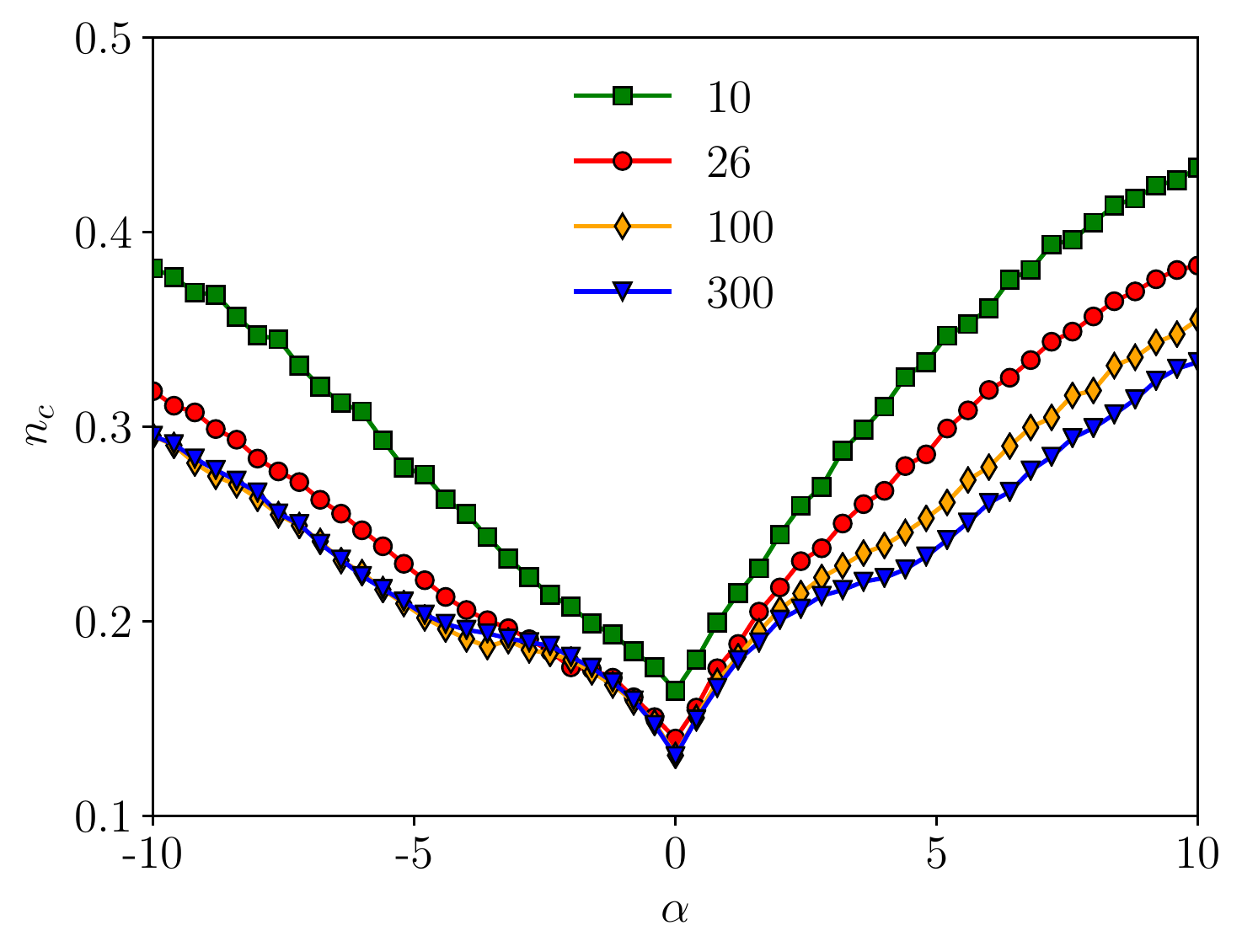}} \\
\subfigure{\includegraphics[width=0.48\textwidth]{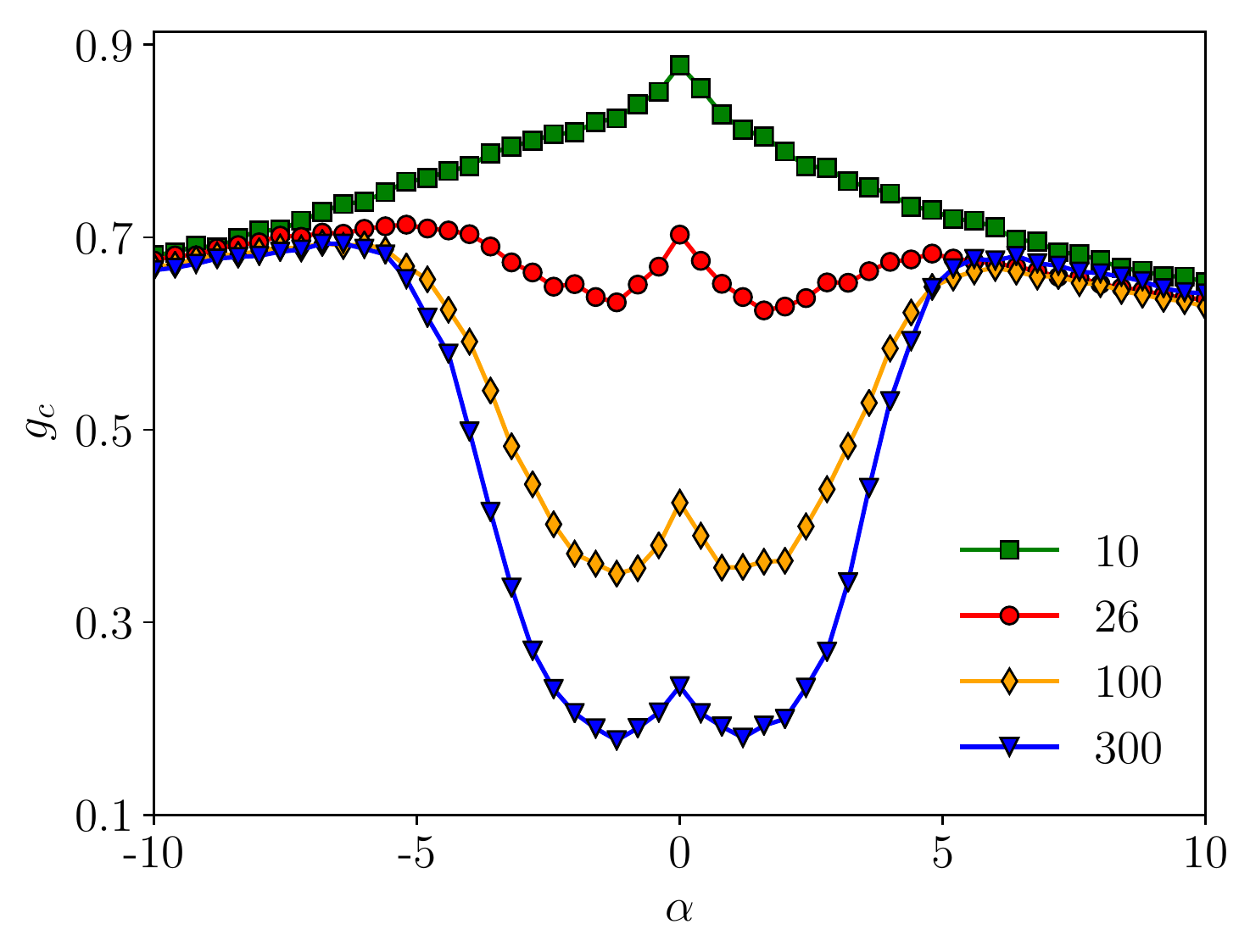}}
\caption{Fraction of strongly connected components $n_c$  (upper panel) and fraction of agents in the largest  strongly connected component $g_c$ (lower panel)  for  group sizes $M=10, 26, 100, 300$ as indicated. 
The imitation probability is $p=0.5$ and the parameters of the NK landscape are $N=12$ and $K=0$.  }
\label{fig:6}
\end{figure}

Since for $\alpha \neq 0$ the sizes of the influence neighborhoods vary as the agents explore the fitness landscape through the imitative search, it is of interest  to characterize the influence networks when the search halts. As pointed out, an influence network, which  is formed by the union of the influence neighborhoods of all agents, is a directed graph where the agents are the nodes and neighboring agents are connected by directed edges  (see Fig.\ \ref{fig:1}).  Since the nub of the imitative search is to spread useful information (i.e., bits that increase fitness) among the members of  the group, we  focus only on  the  connectivity properties  of the influence networks.

We recall that  a directed graph is said to be strongly connected if every node  is reachable from every other node, and a strongly connected component (SCC) of a directed graph is a maximal strongly connected subgraph \cite{Newman_10}. Thus, following the usual line of analysis used to study percolation \cite{Stauffer_92}, we consider the number $\mathcal{N}_c$  of SCCs and  the size  $\mathcal{G}_c$ of  the largest SCC of  the influence network when the search halts. By the size of a SCC we mean the number of nodes that belong to it. Figure \ref{fig:6} shows these two  quantities in a properly normalized form, viz., $n_c = \mathcal{N}_c/M$ and  $g_c=\mathcal{G}_c/M$. The fraction of SCCs reaches its minimum value for  $\alpha =0$  and is  slightly asymmetric around that point, i.e.,  there is a bit more SCCs for positive than for negative $\alpha$.   The size of the largest SCC exhibits this slight asymmetry too. The quasi-invariance of the properties of the SCCs 
to change in the sign of $\alpha$ is expected  since when $\alpha$  is replaced by $-\alpha$ there is essentially an interchange  of the  influence neighborhoods  between  agents  whose fitness  are  above and below the mean fitness of the group by similar amounts.

 The fact that the largest SCC occurs for $\alpha =0$  for small group sizes and contains about 90\% of  the agents  may be the reason that the egalitarian policy is optimal in this situation (see Fig.\ \ref{fig:2}). For large $M$, however, there is no obvious link between $g_c$ and the  computational cost, since $g_c$ is practically the same for $\alpha= 10$ and $\alpha = -10$, but the costs are very distinct (see Fig.\ \ref{fig:3}).

It is interesting to  note that,  for $\mid \alpha \mid < 5$, $g_c$ goes to zero as  $M$ increases   so that the directed graph is composed of a macroscopic number of microscopic SCCs  since $n_c$ is nonzero. For $\mid \alpha \mid >5$ we also have a macroscopic number of SCCs, but now at least one component is macroscopic. In time, a quantity is macroscopic (microscopic) if it grows linearly (sublinearly) with $M$  in the asymptotic limit $M \to \infty$. Moreover, we observe in Fig.\ \ref{fig:6} that there is a region where both  $g_c$ and $n_c$ increase with increasing $\mid \alpha \mid $. This means that while the largest SCC takes in new nodes, other  components break into small components. This is expected considering the dual effect of  increasing $\mid \alpha \mid $ which increases the influence neighborhood of some agents and decrease of others, probably producing isolated agents which would explain the increase of $n_c$.

\subsection{Rugged Landscapes}

\begin{figure}
\centering
\includegraphics[width=0.48\textwidth]{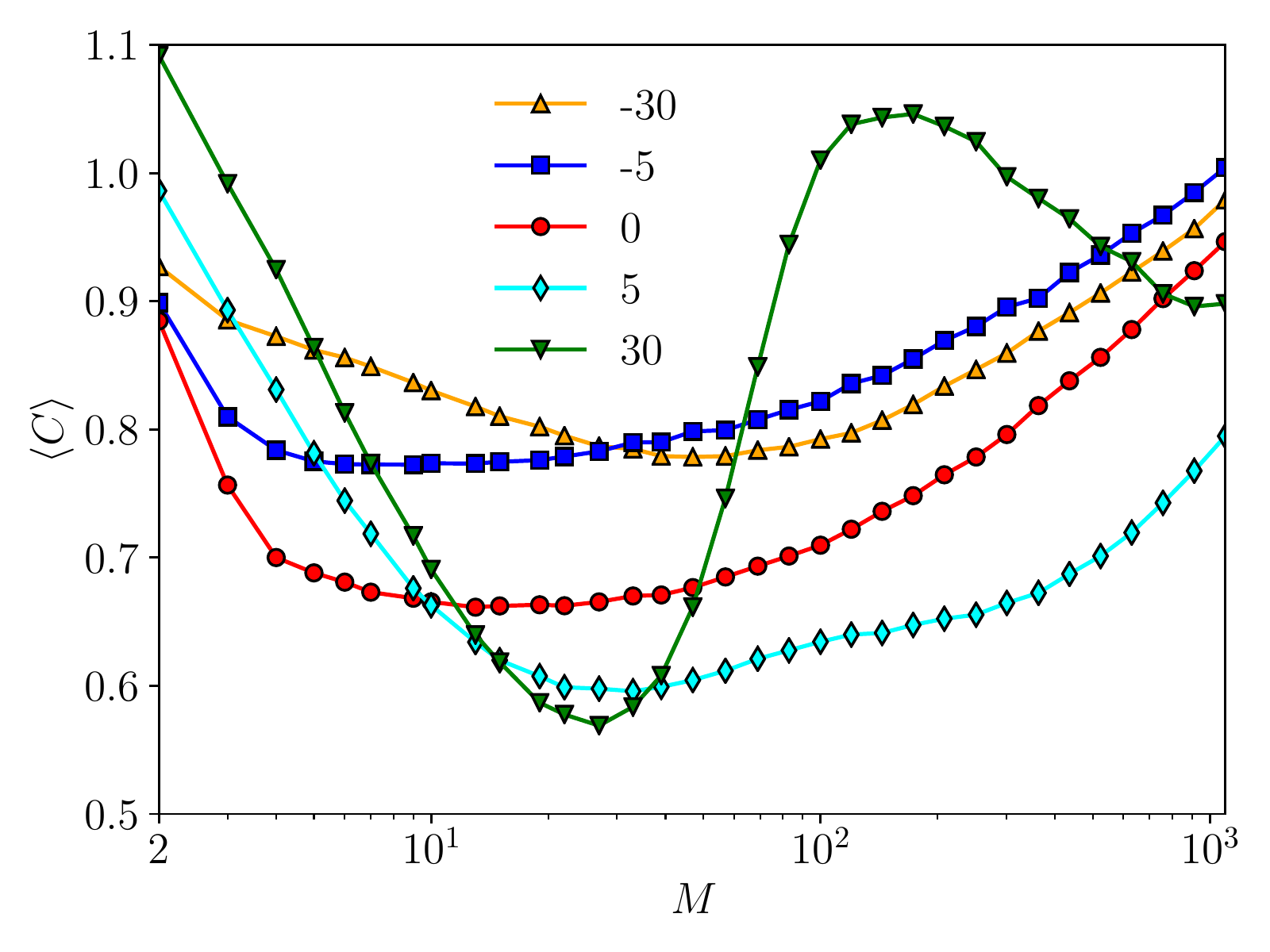}
\caption{Mean computational cost $\langle C \rangle $ as function of the group size $M$ for 
the imitative search on  rugged landscapes. 
The imitation probability is $p=0.5$ and  the strength with which the  information allocation policies are enforced is $\alpha = -30, -5, 0, 5, 30$ as indicated. 
 The parameters of the NK landscapes are $N=12$ and $K=4$.  }
\label{fig:1K4}
\end{figure}

\begin{figure}
\centering
\includegraphics[width=0.48\textwidth]{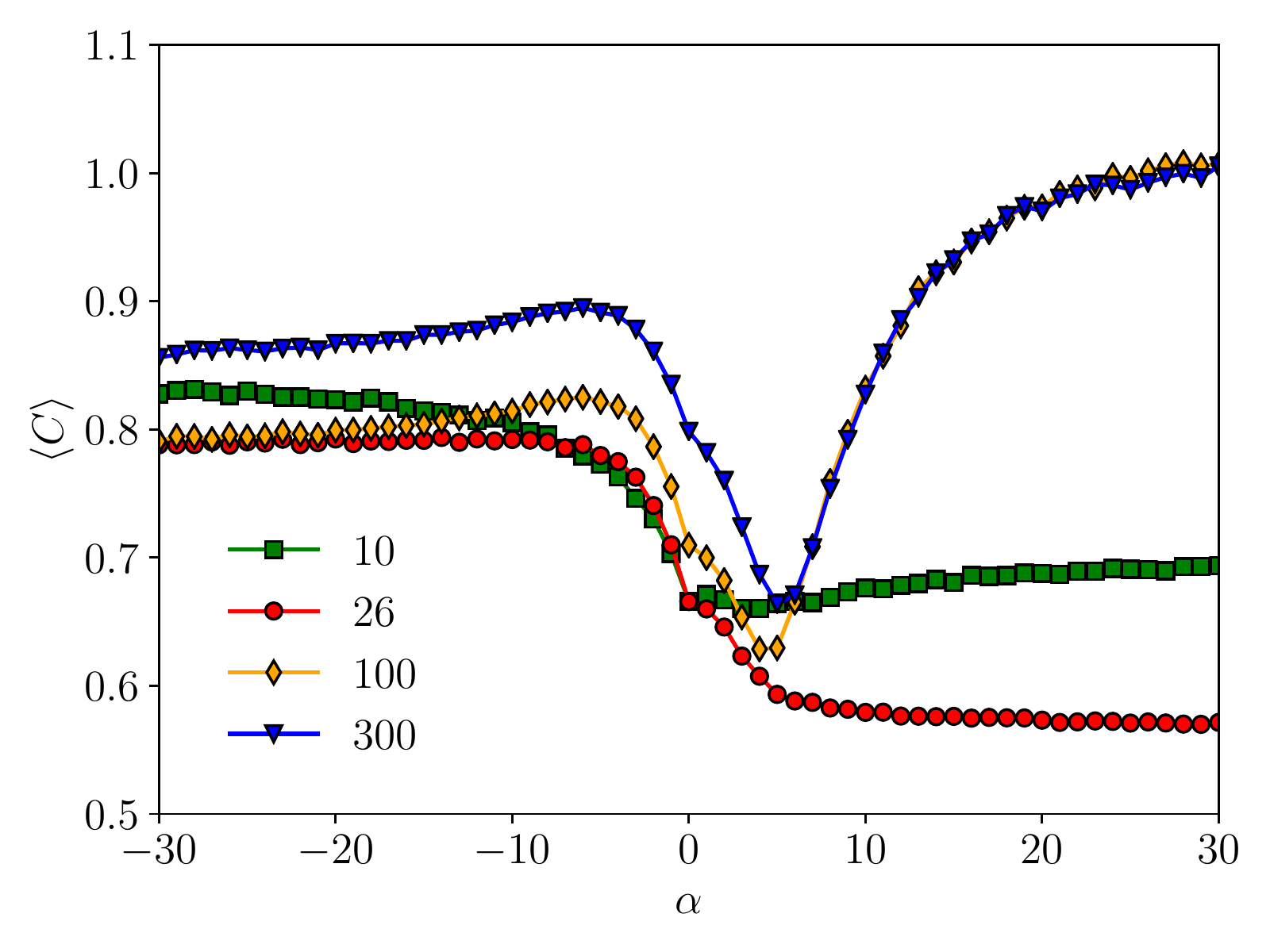}
\caption{Mean computational cost $\langle C \rangle $ as function of the strength $\alpha$  with which the  information allocation policies are enforced for   group sizes $M=10,26, 100, 300$ as indicated. 
The imitation probability is $p=0.5$  and the parameters of the NK landscapes are $N=12$ and $K=4$.  }
\label{fig:2K4}
\end{figure}

Since even the less rugged landscape in our ensemble  of 30 NK-fitness  landscapes  with parameters  $N=12$ and $K=4$ has 31 maxima,   finding the unique global maximum of each realization in this set poses  a difficult challenge to any hill-climbing type of search strategy. In fact, the presence of those local maxima makes the computational cost of the imitative search very susceptible  to the choice of the group size  and of the policy to allot  information to the agents.  Figures \ref{fig:1K4} and \ref{fig:2K4} illustrate the intricacies of this choice. 

%
%

Whereas  the best performance shown in Fig. \ref{fig:1K4} is a\-chieved by the highly elitist policy ($\alpha = 30$) for $M \approx 26$, this policy gives the  worst performance for groups of small and intermediate  sizes. In particular, the peak  of the cost observed  for  groups of 
intermediate size  can be viewed as  a groupthink-like phenomenon \cite{Janis_82} that happens because the agents are trapped in high fitness local maxima far away from the global maximum. The cost to escape those maxima can be very high  due to  the attractive effect of the clones of the model string \cite{Fontanari_15}.  Nevertheless, the computational cost of the imitative search is  always lower than the cost of the independent search, which is  $\langle C \rangle \approx 1.1$ for the range of $M$ shown in the figure. This contrasts with the results for the fully connected system for which 
the maximum  cost  of the imitative search is much higher than the cost of the independent search \cite{Fontanari_15}. The reason is that the prescription (\ref{dk}) for the radiuses of the influence neighborhoods of the agents introduces an effective diversity in the imitation probabilities of the agents. In fact, since   isolated agents can only flip bits at random  their effective probability of imitation is zero. It is the presence of those agents in the group that prevents the trapping of the entire group in a local maximum, which would  then characterize a  full-blown groupthink event \cite{Fontanari_16}.

Interestingly, for small group sizes, the best performance is achieved by the egalitarian policy ($\alpha = 0$) as in the case of the smooth landscape. In fact,  for small group sizes the performances of the distinct allocation policies is little influenced by the ruggedness of the landscapes (see Figs.\ \ref{fig:2} and \ref{fig:1K4}).

Figure  \ref{fig:2K4} shows that the welfarist policy ($\alpha < 0$) is always suboptimal, regardless of  the value of $M$. By suboptimal we mean that either $\alpha=0$ or $\alpha >0$  yield a better performance for a fixed $M$. This conclusion holds true for smooth and rugged landscapes as well (see Figs.\ \ref{fig:3} and  \ref{fig:2K4}).

%
%

\begin{figure}
\centering
\subfigure{\includegraphics[width=0.48\textwidth]{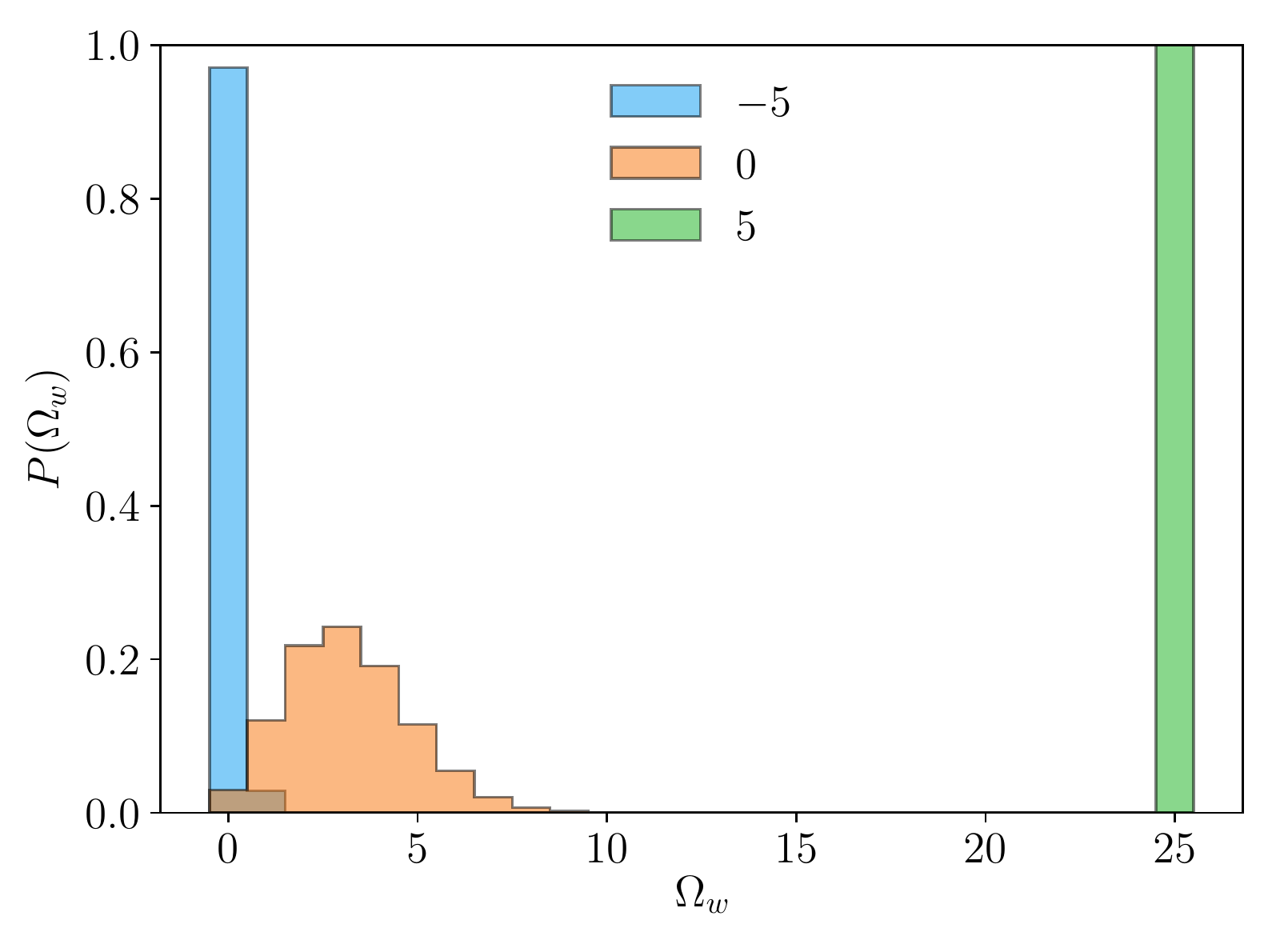}} \\
\subfigure{\includegraphics[width=0.48\textwidth]{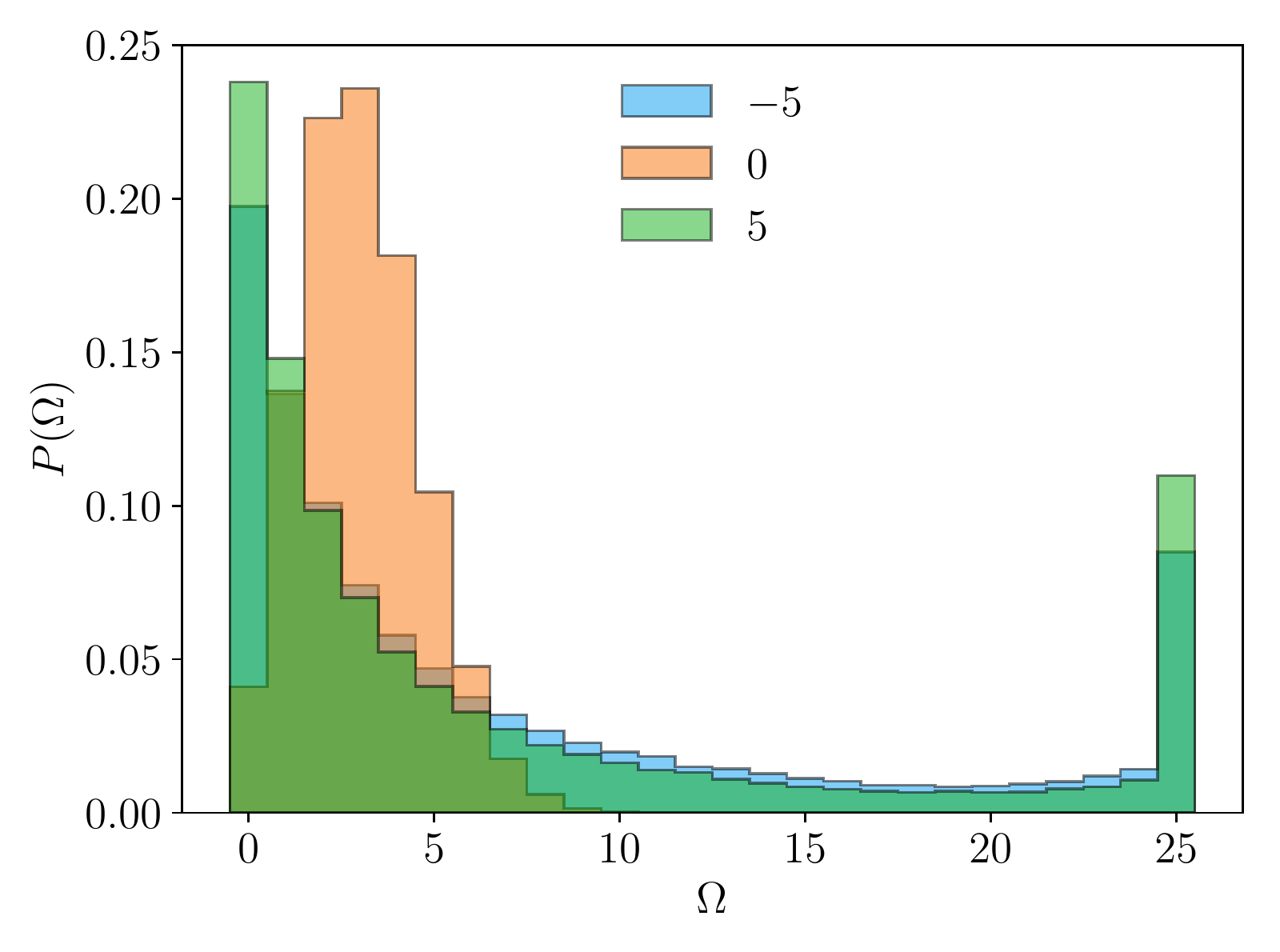}}
\caption{Probability distribution of the number of agents in the influence neighborhood of the winner (upper panel) and of 
a randomly selected agent (lower panel)  at the  instant just before the winner finds the global maximum for 
 $M=26$ and $\alpha =-5,0,5$   as indicated. 
The imitation probability is $p=0.5$ and the parameters of the NK landscapes are $N=12$ and $K=4$. }
\label{fig:3K4}
\end{figure}

\begin{figure}
\centering
\includegraphics[width=0.48\textwidth]{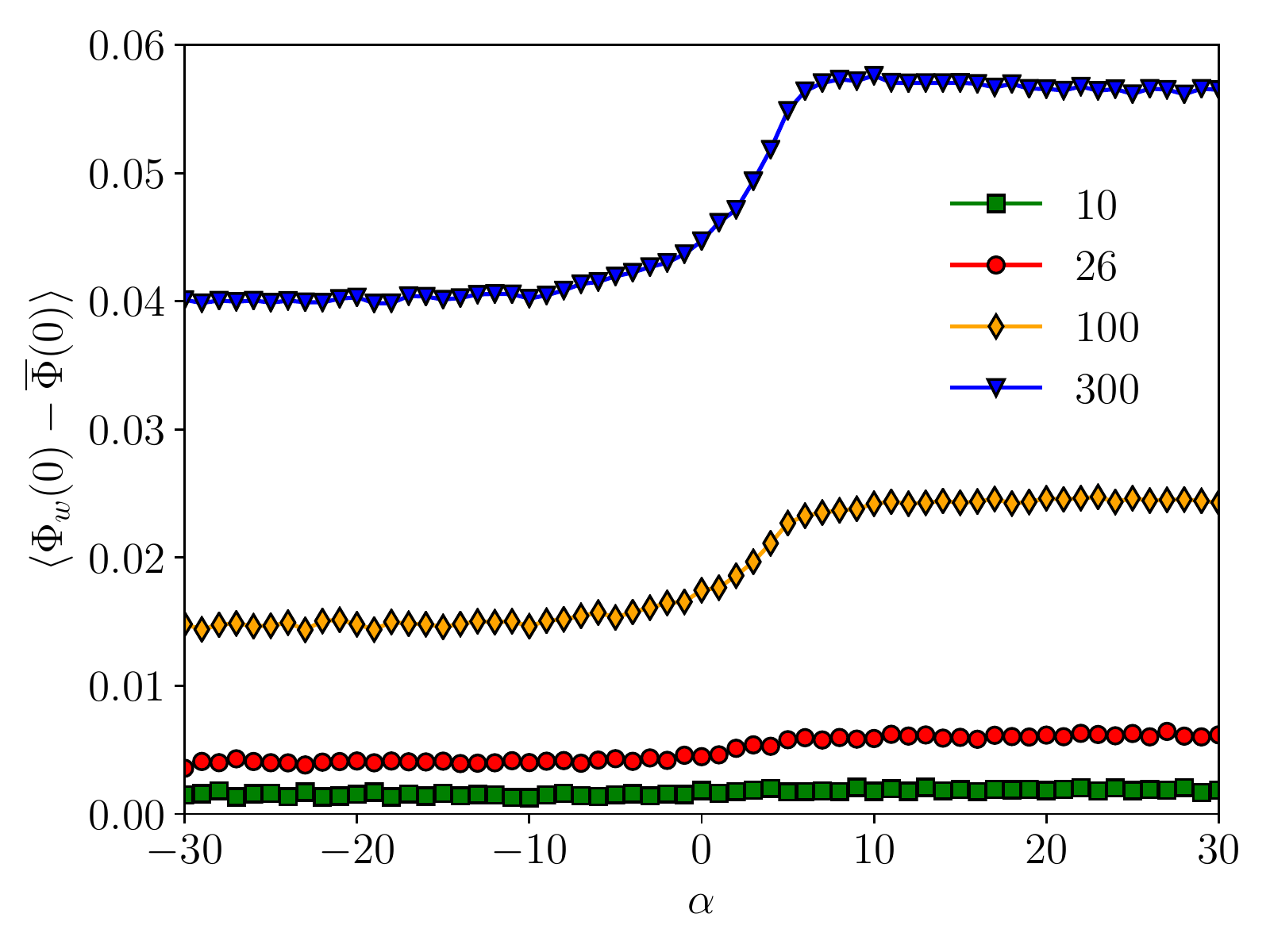}
\caption{Average difference between the initial fitness of  winner   and the  initial mean fitness of the group 
$\left  \langle \Phi_w \left ( 0 \right ) - \bar{\Phi} \left ( 0 \right ) \right \rangle$ as function  of
the strength $\alpha$  with which the  information allocation policies are enforced  for  group sizes $M=10,26, 100, 300$ as indicated. 
The imitation probability is $p=0.5$  and the parameters of the NK landscapes are $N=12$ and $K=4$.  }
\label{fig:4K4}
\end{figure}

Figure \ref{fig:3K4} shows the distribution of 
the number of agents in the influence neighborhood  of the winner and of a randomly selected agent at the instant just before the winner finds the global maximum. The winner is almost certainly isolated for $\alpha < 0$ or  fully connected for $\alpha > 0$, although  the size of the influence neighborhood of a randomly selected agent is very little affected by the sign of $\alpha$.  Use of the elitist policy ($\alpha > 0$)   gives an advantage to high-fitness outliers produced in the initial setup of the group as shown in  Fig.\ \ref{fig:4K4}, but it is not as significant as in the case of the smooth landscape (see Fig.\ \ref{fig:5}).
 This is probably because enlarging the influence neighborhood of an agent is not necessarily advantageous for the group performance as the fitness of the model agents are not strongly correlated to the distances to the global maximum
as happens for the smooth landscape.

To conclude, we note that the quantities $n_c$ and $g_c$ for the rugged landscapes are very similar to those shown in Fig.\ \ref{fig:6} for the smooth landscape. Hence the difficulty of the problem posed to the group does not seem to influence the connectivity properties of the influence networks when the search halts.

 \section{Discussion}\label{sec:conc}
 
Problem solving by small or large groups of people  is a critical issue in modern life  as evinced  by today's  highly successful web-enabled collective intelligence enterprises such as Google and Wikipedia \cite{Malone_10}.  Hence the relevance of understanding the factors that influence the capability  of
  task-oriented groups  to solve problems.   Here we approach this issue by combining ideas  from  organizational  design \cite{Lazer_07} and the theory of   distributed cooperative problem-solving  systems  \cite{Huberman_90}.

 It has long been realized that  the patterns of communication  that determine who can communicate  with whom in a task-oriented group  have a great impact on its  problem-solving performance  both in the case where cooperation is mandatory to solve the task (e.g.,  finding the common card in  decks distributed to  subjects  as in Leavitt-Bavelas' experiment \cite{Bavelas_50,Leavitt_51,Reia_19a})
and  in the case where a single individual could in principle solve the task (e.g., finding the global maximum of a fitness landscape  \cite{Mason_12,Sebastian_17}). However, these studies have focused on  imposed or fixed  communication patterns,  thus excluding a priori  the interesting possibility  of self-organization of the communication networks. 

 Here we explore a  scenario of flexible  communication patterns where  immobile  agents vary their radiuses of interaction according to the (relative) quality of the  solutions they offer to the  problem posed to the group, which is to find the  global maximum of a NK-fitness landscape. The group performance is measured by a computational cost that essentially tallies the total number of bit flips  performed on the $M$  binary strings (i.e., agents)  that compose the group  until the global maximum is found.
 
 The variation of the sizes of the  influence neighborhoods  of the agents results in a time-dependent,  adaptive directed network that links the agents to their influencers.  Since the size of  the influence neighborhood of an agent is a measure of the amount of information it can use to decide which bit to flip, it is  necessary to establish a policy for allocation of
information to the agents based on  their fitness. Here we consider three information allocation strategies that are determined by the sign of the parameter $\alpha$ in eq. (\ref{dk}). The first is the elitist policy in which  agents with above-average fitness  have their influence neighborhoods amplified, whereas agents  with below-average fitness  have theirs deflated.  The second is the welfarist policy in which the actions of the elitist policy are reversed, and the third is the egalitarian policy in which  the size of the influence neighborhood is the same for all agents. 
 
Policies for allocation of information  are of great importance when the links or connections between individuals are costly, as in the case of  social networks of gregarious animals 
where there is a direct selection pressure to reduce  the number of connections between entities because of their  building and maintenance costs \cite{Waters_12,Pasquaretta_14,Kurvers_14}. In addition, the view of science as a massive, real-world collective search problem and of
scientists as  single solution-searching units  \cite{Kitcher_93,Goldstone_08} brings forth the issue of how to allot resources to  competing scientists. Resources that are typically used for `networking' as in the scenario described here. Moreover, the notions of exploration (discovering new results) and exploitation (borrowing results from others) are not strange to the scientific  enterprise.
 In this context, a natural criterion to allot funds to  scientists  is their  reputations.  
In the  context of searching for the global optimum of a NK-fitness landscape, the relative fitness plays the role of the scientist's reputation,  hence our proposal of  the prescription (\ref{dk}) to define the radiuses of the influence neighborhoods of the agents.

Somewhat surprising, we find that for small groups the egalitarian policy is optimal for both smooth and rugged landscapes
(Figs.\ \ref{fig:2} and \ref{fig:1K4}).  In addition, we find that the elitist policy is   optimal for smooth landscapes, provided the group size is not too small. However, this policy produces disastrous results for  groups of intermediate sizes in the case of rugged landscapes, which is akin to the groupthink phenomenon of social psychology \cite{Janis_82} that results from the lack of opinion diversity   among the group members. The welfarist policy, on the other hand, is always suboptimal and, in particular, it is always outperformed by the egalitarian policy. 

An interesting and realistic addition to our problem solving-solving scenario is to consider that the  members within the group  are subject to a social network besides the professional network studied here. The fixed-topology social network could  be  considered as a second layer of a two-layer network \cite{Dickison_16}, with the first layer being our adaptive influence network. On the one hand, this scenario will prevent the appearance of isolated agents, which could improve the group performance. On the other hand, the effective increase of the network connectivity  due to the extra layer may degrade the group performance by magnifying the groupthink phenomenon   \cite{Francisco_16}. This tradeoff makes the study of the two-layer network scenario an attractive  research program.

An appealing finding about the elitist policy regards its potential to select high-fitness outliers in the initial randomly generated group  as the  winner of the search, i.e., the agent that finds the global maximum first. For both  smooth and rugged  landscapes,  the  elitist policy picks winners with a much higher initial mean  fitness  than those of  the other two policies (Figs.\ \ref{fig:5} and \ref{fig:4K4}), in accordance with the Matthew principle that `the rich get richer and the poor get poorer'. 
More interesting, however,  is the finding that even in a situation of strong welfare, say $\alpha = -30$ in those figures,  where the high-fitness agents are isolated and the low-fitness agents  are allowed access to the entire group, the high-fitness outliers of the initial generation are  still more likely to become winners, so our welfarist  policy cannot reverse the random initial fitness inequality. These conclusions are valid for large groups only, for which  there is a reasonable chance  of producing a random string  with fitness much higher than the average of the group. 

The characterization of the influence networks using the distribution of the sizes of the influence neighborhoods (lower panels of Figs. \ref{fig:4} and \ref{fig:3K4}) and the statistics of the strongly connected components  (Fig.\ \ref{fig:6})  reveal the rich  topology produced  by the interplay between the network structure and the imitative search dynamics. We conclude that, except for small groups, some degree of flexibility on the communication  patterns  among agents can be beneficial  to the  group performance, provided an elitist policy is enforced with moderate   strength.

\bigskip

\acknowledgments
The research of JFF was  supported in part 
 by Grant No.\  2017/23288-0, Fun\-da\-\c{c}\~ao de Amparo \`a Pesquisa do Estado de S\~ao Paulo 
(FAPESP) and  by Grant No.\ 305058/2017-7, Conselho Nacional de Desenvolvimento 
Cient\'{\i}\-fi\-co e Tecnol\'ogico (CNPq).
SMR  was supported by grant  	15/17277-0, Fun\-da\-\c{c}\~ao de Amparo \`a Pesquisa do Estado de S\~ao Paulo 
(FAPESP).

\end{document}